# PERFORMANCE ANALYSIS OF NEURAL NETWORK MODELS FOR OXAZOLINES AND OXAZOLES DERIVATIVES DESCRIPTOR DATASET

Doreswamy and Chanabasayya .M. Vastrad

Department of Computer Science Mangalore University, Mangalagangotri-574 199, Karnataka, India

## ABSTRACT

*Neural networks have been used successfully to a broad range of areas such as business, data mining, drug discovery and biology. In medicine, neural networks have been applied widely in medical diagnosis, detection and evaluation of new drugs and treatment cost estimation. In addition, neural networks have begin practice in data mining strategies for the aim of prediction, knowledge discovery. This paper will present the application of neural networks for the prediction and analysis of antitubercular activity of Oxazolines and Oxazoles derivatives. This study presents techniques based on the development of Single hidden layer neural network (SHLFFNN), Gradient Descent Back propagation neural network (GDBPNN), Gradient Descent Back propagation with momentum neural network (GDBPMNN), Back propagation with Weight decay neural network (BPWDNN) and Quantile regression neural network (QRNN) of artificial neural network (ANN) models Here, we comparatively evaluate the performance of five neural network techniques. The evaluation of the efficiency of each model by ways of benchmark experiments is an accepted application. Cross-validation and resampling techniques are commonly used to derive point estimates of the performances which are compared to identify methods with good properties. Predictive accuracy was evaluated using the root mean squared error (RMSE), Coefficient determination( $R^2$ ), mean absolute error(MAE), mean percentage error(MPE) and relative square error(RSE). We found that all five neural network models were able to produce feasible models. QRNN model is outperforms with all statistical tests amongst other four models.*

## 1. INTRODUCTION

The use of artificial neural networks (ANNs) in the area of drug discovery and optimization of the dosage forms has become a topic of analysis in the pharmaceutical literature [1-5]. Compared with linear modelling techniques, such as Multi linear regression (MLR) and Partial least squares (PLS), ANNs show better as a modelling technique for molecular descriptor data sets showing non-linear conjunction, and thus for both data fitting and prediction strengths [6]. Artificial neural network (ANN) is a vastly simplified model of the form of a biological network[7] .The fundamental processing element of ANN is an artificial neuron (or commonly a neuron). A

DOI : 10.5121/ijist.2013.3601            1

biological neuron accepts inputs from other sources, integrate them, carry out generally a nonlinear process on the result, and then outputs the last result [8]. The fundamental benefit of ANN is that it does not use any mathematical model because ANN learns from data sets and identifies patterns in a sequence of input and output data without any previous assumptions about their type and interrelations [7]. ANN eliminates the drawbacks of the classical ways by extracting the wanted information using the input data. Executing ANN to a system uses enough input and output data in place of a mathematical equation[9]. ANN is a good alternative to common empirical modelling based on linear regressions [10].

ANNs are known to be a powerful methodology to simulate various non-linear systems and have been applied to numerous applications of large complexity in many field including pharmaceutical research, engineering and medicinal chemistry. The promising uses of ANN approaches in the pharmaceutical sciences are widespread. ANNs were also widely used in drug discovery, especially in QSAR studies. QSAR is a mathematical connection between the chemical's quantitative molecular descriptors and its inhibitory activities [11-12]

Five Different types of neural network models have been developed for the development of efficient antitubercular activity predicting models. Those models are Single hidden layer feed forward neural network (SHLFNN)[13],Gradient Descent Back propagation neural network (GDBPNN)[14,24], Gradient Descent Back propagation with momentum neural network (GDBPMNN)[15-16],Back propagation with Weight decay neural network (BPWDNN)[17],Quantile regression neural network (QRNN) [18].

The purpose of this research work and research publication is to assign five distinct neural network models to the prediction of antitubercular activities of Oxazolines and Oxazoles derivatives descriptor dataset. Method and along with estimate and asses their performances with regard to their predicting ability. One of the goals of this scientific research project is to show how distinct neural network models can be used in predicting antitubercular activities of Oxazolines and Oxazoles derivatives descriptor dataset. It again involves inducing the best model in terms of the least errors produced in the graphical study describing the actual and predicted antitubercular activities.

## 2. MATERIALS AND ALGORITHAMS

### 2.1 The Data Set

The molecular descriptors of 100 Oxazolines and Oxazoles derivatives [19-20] based H37Rv inhibitors analyzed. These molecular descriptors are produced using Padel-Descriptor tool [21]. The dataset includes a different set of molecular descriptors with a broad range of inhibitory activities versus H37Rv. This molecular descriptor data set includes 100 observations with 234 descriptors. Before modelling, the dataset is ranged.

### 2.2 Single hidden layer feed forward neural network (SHLFNN)

The clearest form of neural network is one among a single input layer and an output layer of nodes. The network in Figure 1 represents this type of neural network. Strictly, this is mentioned to as a one-layer feed forward network among two outputs on account of the output layer is the alone layer with an activation computation.

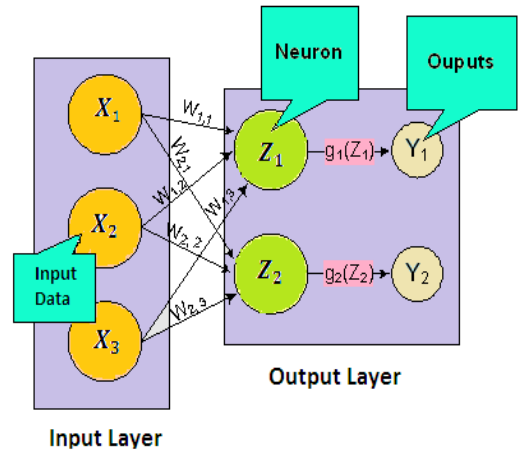

Figure 1. A Single Hidden Layer Feed Forward Neural Network

In this single hidden layer feed forward neural network, the network's inputs are directly connected to the output layer nodes, $Z_1$ and $Z_2$. The output nodes use activation functions $g_1$ and $g_2$ to yield the outputs $Y_1$ and $Y_2$.

Because

$$Z_1 = \sum_{i=1}^{3} W_{1,i} x_i + \mu_1 \quad and \quad Z_2 = \sum_{i=1}^{3} W_{2,i} x_i + \mu_2$$

$$Y_1 = g_1(Z_1) = g_1\left(\sum_{i=1}^{3} W_{1,i} x_i + \mu_1\right) \tag{1}$$

and

$$Y_2 = g_2(Z_2) = g_2\left(\sum_{i=1}^{3} W_{2,i} x_i + \mu_2\right) \tag{2}$$

When the activation functions $g_1$ and $g_2$ are similar activation functions, the single hidden layer feed forward neural network is similar to a linear regression model. Likewise, if $g_1$ and $g_2$ are logistic activation functions, then the single hidden layer feed forward neural network is similar to logistic regression. Because of this comparison between single hidden layer feed forward neural networks and linear and logistic regression, single hidden layer feed forward neural networks are not often used in place of linear and logistic regression.

## 2.3 Gradient Descent Back Propagation Neural Network(GDBPNN)

Gradient Descent Back propagation neural network is one of the most engaged ANN algorithms in pharmaceutical research. GDBPNN are the nearly general type of feed-forward networks. Figure 2 displays an back propagation neural network which has three types of layers: an input layer, an output layer and a hidden layers.

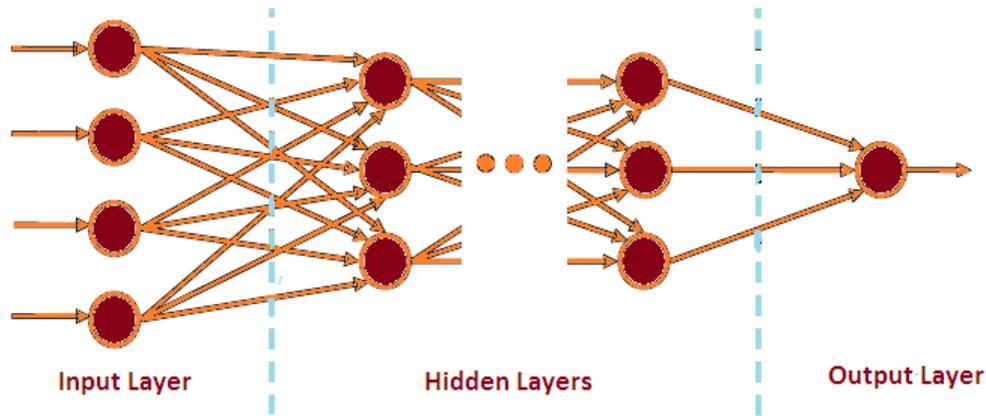

Figure 2. A Back propagation (BP) neural network

Nodes(neurons) in input layer only act as buffers for delivering the input data $x_i$ ($i = 1,2 \ldots n$) to nodes in the hidden layer. Each processing node $j$ (Figure 3) in the hidden layer sums up its input data $x_i$ after weighting them with the strengths of the particular connections $w_{ji}$ from the input layer and calculates its output $y_j$ as a function $f$ of the sum.

$$y_j = f(\sum_{i=1}^{n} w_{ji} x_i) \qquad (3)$$

Activation function $f$ that is generally selected to be the sigmoid function.

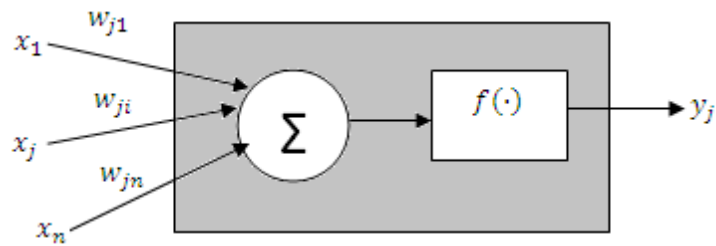

Figure 3. Specification of the perceptron process

The output of nodes in the output layer is calculated similarly. The backpropagation gradient descent algorithm, is the most generally approved Multi Layer Perciptron training algorithm. It provides to alter $\Delta w_{ji}$ the weight of a connection between nodes i and j as accordingly:

$$\Delta w_{ji} = \eta \delta_j x_i \qquad (4)$$

Where $\eta$ is a parameter termed the learning rate and $\delta_j$ is a factor depending on whether node $j$ as an input node or a hidden node. For output nodes,

$$\delta_j = (\partial f / \partial \, nnet_j)\left(y_j^{(t)} - y_j\right) \qquad (5)$$

and for hidden nodes

$$\delta_j = (\partial f / \partial\, nnet_j)\left(\sum_q w_{jq}\delta_q\right) \qquad (6)$$

In Eq. (5), $nnet_j$ is the aggregate weighted sum of input data to nodes $j$ and $y_j^{(t)}$ is the target output for node $j$. As there are no target outputs for hidden nodes, in Eq. (6), the variation between the target and measured output of a hidden nodes $j$ is put back by the weighted aggregate of the $\delta_q$ terms at present obtained for nodes $q$ linked to the output of node $j$ The method starts with the output layer, the $\delta$ term is calculated for nodes in entire layers and weight updates detected for all links, repetitively. The weight updating method can happen after the presentation of each training observation (observation-based training) or after the presentation of the whole set of training observations. Training epoch is achieved when all training patterns have been introduced once to the Multilayer Perceptron.

## 2.4 Gradient Descent Back Propagation with Momentum Neural Network (GDBPMNN)

Gradient descent back propagation with momentum neural network (GDBPMNN) algorithm is widely used in neural network training, and its convergence is discussed. A momentum term is often added to the GDBPNN algorithm in order to accelerate and stabilize the learning procedure in which the present weight updating increment is a mixture of the current gradient of the error function and the prior weight revising increment. Gradient decent back propagation with momentum allows a neural network to respond not only to the local gradient, but also to recent tendency in the error surface. Momentum allows the neural network to ignore small features in the error surface. Without momentum a neural network may get stranded in a shallow local minimum. With momentum a network can move through such a least.

Momentum can be combined to GDBPNN method learning by building weight alters balance to the sum of a portion of the final weight modification and the new modification advised by the GDBP rule. The importance of the response that the last weight modification is admitted to have is negotiated by a momentum constant, $\mu$, which can be any number between 0 and 1. When the momentum constant is 0 a weight modification is based only on the gradient. When the momentum constant is 1 the new weight modification is set to balance the last weight modification and the gradient is plainly neglected.

$$\Delta w_{ij}(I + 1) = \eta \delta_j x_i + \mu \Delta w_{ij}(I) \qquad (7)$$

where $\Delta w_{ij}(I + 1)$ and $\Delta w_{ij}(I)$ are weight alters in epochs $(I + 1)$ and $(I)$, suitable way[24].

## 2.5 Back propagation with Weight Decay Neural Network (BPWDNN)

Back propagation of error gradients for back propagation neural networks has proven to be useful in layered feed forward neural network training. Still, a wide number of repetitions is commonly required for changing the weights. The problem becomes more critical especially when a high level of accuracy is required. The complexity of a back propagation neural network can be regulated by a hyper-parameter called "weight decay" to penalize the weights of hidden nodes.

International Journal of Information Sciences and Techniques (IJIST) Vol.3, No.6, November 2013The employ of weight decay can both assist the optimization deals with and prevent the over fitting. This type of method to encourage the learning algorithm to find solutions which use as few weights as possible. The simplest modified error function can be formed by summing to the initial error function a term relative to the sum of squares of weights:

$$E = E_i + \lambda \sum_i \sum_j w_{ij}^2 \tag{8}$$

where $E_i$ is the initial error function (sum of the squared differences between actual and predicted output values), $\lambda$ is a minute positive constant which is employed to govern the addition of the second term, and $w_{ij}$ is the weight of the link between node j and of a layer and node i of the at once higher indexed layer. The above error function penalizes the use of more $w_{ij}$'s than essential. In order to demonstrate that, lets see how the weight updating rule is changed. Assuming that we apply the gradient descent algorithm to minimize the error, the changed weight update method is shown by:

$$\Delta w_{ij}(I) = -\eta \left(\frac{\partial E}{\partial w_{ij}}\right)(I) = -\eta \left(\frac{\partial E_i}{\partial w_{ij}}\right)(I) - 2\lambda \eta w_{ij}(I) \tag{9}$$

where $I$ denotes the $I$-th iteration and $\eta$ denotes the learning rate. The above expression can be composed as

$$w_{ij}(I+1) = -\eta \left(\frac{\partial E_i}{\partial w_{ij}}\right)(I) + (1 - 2\lambda\eta) w_{ij}(I) \tag{10}$$

It can be demonstrates that the importance of the weights decreases exponentially towards zero by calculating the weight values after $I$ weight adaptations:

$$w_{ij}(I) = \eta \sum_{i=1}^{I} (1 - 2\lambda\eta)^{I-i} \left(-\left(\frac{\partial E_i}{\partial w_{ij}}\right)\right) + (1 - 2\lambda\eta)^I w_{ij}(0) \tag{11}$$

(assuming $|1 - 2\lambda\eta| < 1$). The above method has the disadvantage that all the weights of the neural network decrease at the same rate. Still, it is more attractive to allow large weights to carry on while small weights tend toward zero. This can be carried out by modifying the error function in a way that small weights are altered more considerably than large weights. The following modified function:

$$E = E_i + \lambda \sum_i \sum_j \frac{w_{ij}^2}{1 + w_{ij}^2} \tag{12}$$

The weight updating rule then becomes:

$$w_{ij}(I+1) = -\lambda\left(\frac{\partial E_i}{\partial w_{ij}}\right)(I) + \left(1 - \frac{2\lambda\eta}{\left(1+w_{ij}^2(I)\right)^2}w_{ij}(I)\right) \quad (13)$$

It can be demonstrates that in this case small weights decrease more swiftly than large ones.

## 2.6 Quantile Regression Neural Network (QRNN)

Artificial neural networks allow the estimation of in some way nonlinear models without the need to define a accurate functional form. The most widely-used neural network for predicting is the single hidden layer feed forward neural network [25]. It exists a set of n input nodes , which are connected to each of m nodes in a single hidden layer, which, in turn, are connected to an output node. The final model can be made as

$$f(x_t, v, w) = g_2\left(\sum_{j=0}^{m} v_j\, g_1\left(\sum_{i=0}^{n} w_{ji}\, x_{it}\right)\right) \quad (14)$$

where $g_1(\cdot)$ and $g_2(\cdot)$ are activation functions, which are commonly chosen as sigmoidal and linear accordingly, $w_{ji}$ and $v_j$ are the weights to be approximated.

Theoretical assist for the use of quantile regression within an artificial neural network for the evaluation of probably nonlinear quantile models [26]. The only other work that we are knowledgeable of, that considers quantile regression neural networks , is that of Burgess [27] who briefly explains the proposal of the method. Alternative way of linear quantile function using the equation in (14) , a quantile regression neural network model, $f(x_t, v, w)$, of the $\theta th$ quantile can be estimated using the following minimisation.

$$\min_{v,w}\left(\sum_{t|y_t \geq f(x_t,v,w)} \theta|y_t - f(x_t,v,w)| + \sum_{t|y_t \geq f(x_t,v,w)} (1-\theta)|y_t - f(x_t,v,w)| + \lambda_1 \sum_{j,i} w_{ji}^2 + \lambda_2 \sum_i v_i^2\right) \quad (15)$$

where $\lambda_1$ and $\lambda_2$ are regularisation parameters which penalise the complicatedness of the neural network and thus prevent over fitting [28].

## 2.7 Fitting and comparing models

The solutions for the SHLFFNN , GDBPNN, GDBPMNN, BPWDNN and QRNN models were computed using open source CRAN R packages nnet ,neuralnet, RSSNS and qrnn. These five neural network models are trained on a Oxazolines and Oxazoles derivatives descriptor dataset , it constructs a predictive model that returns a minimization in error when the neural network's prediction (its output) is compared with a known or expected outcome. The comparison between the five models were assessed using root mean square error (RMSE) and coefficient of determination $R^2$. RMSE presents information on the short term efficiency which is a benchmark of the difference of predicated values about the observed values. The lower the RMSE, the more accurate is the evaluation and coefficient of determination (also called R square) measures the variance that is interpreted by the model, which is the reduction of variance when using the model. $R^2$ orders from 0 to 1 while the model has healthy predictive ability when it is near to 1 and is not analyzing whatever when it is near to 0. These performance metrics are a good measures of the overall predictive accuracy.

MAE(mean absolute error) is an indication of the average deviation of the predicted values from the corresponding observed values and can present information on long term performance of the models; the lower MAE the better is the long term model prediction. Relative squared error (RSE) is the aggregate squared error produce relative to what the error would have been if the prediction had been the average of the absolute value. Lower RSE is the better model prediction. The Mean Percent Error (MPE) is a well known measure that corrects the 'cancelling out' results and also keeps into basis the different scales at which this measure can be calculated and thus can be used to analyze different predictions. The expressions of all measures are given below.

$$MAE = \frac{1}{n}\sum_{i=1}^{n}|\hat{y}_i - y_i| \qquad (16)$$

$$MPE = \frac{1}{n}\sum_{i=1}^{n}\frac{y_i - \hat{y}_i}{y_i} * 100 \qquad (17)$$

$$RSE = \frac{\sum_{i=1}^{n}(\hat{y}_i - y_i)^2}{\sum_{i=1}^{n}(mean(y) - y_i)^2} \qquad (18)$$

where $y_i$ and $\hat{y}_i$ are observed and predicted values.

### 2.9 Benchmark Experiments

Move in benchmark experiments for comparison of neural network models. The experimental performance distributions of a set of neural network models are estimated, compared, and ordered. The resampling process used in these experiments must be investigate in further detail to determine which method produces the most accurate analysis of model influence. Resampling methods to be compared include cross-validation [29-31]. We can use resampling results to make orderly and in orderly comparisons between models [29-30]    Each model performs 25 independent runs on each sub sample and report minimum, median, maximum, mean of each performance measure over the 25 runs.

## 3. RESULTS AND DISCUSSION

This part presents the numerical analysis conducted using numerous neural network methods. RMSE and $R^2$ values were used to analyze model prediction accuracies for the SHLFFNN,GDBPNN, GDBPMNN, BPWDNN and QRNN neural network models. Comparing the resampling performance the effect of prediction of antituberculer activity using Oxazolines and Oxazoles derivatives are demonstrated in Figure 4.

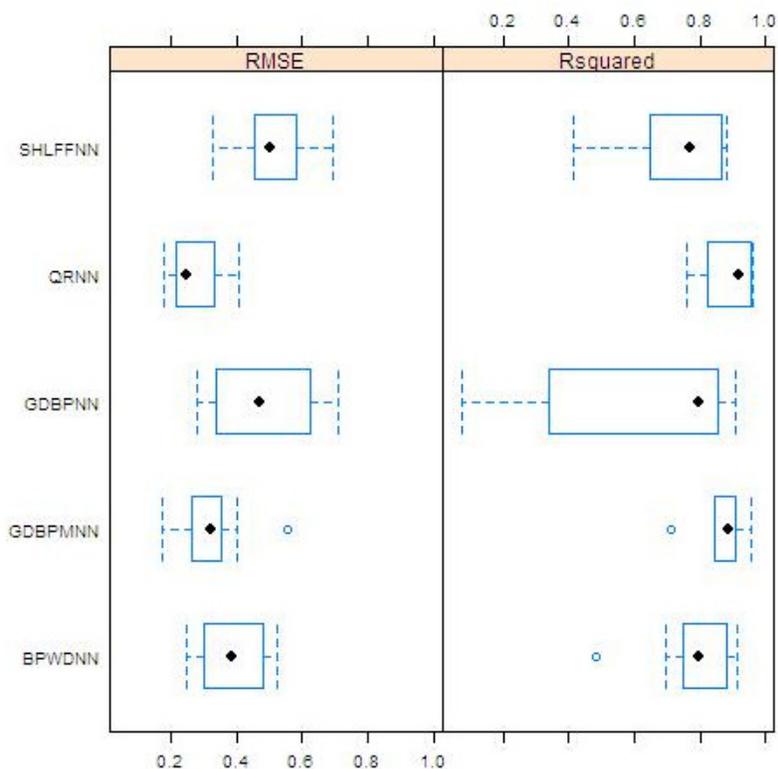

Figure 4. Box-and-whisker diagrams for the cross validation estimates of model precision performance RMSE, or $R^2$. The QRNN and GDBPMNN models gives the smallest prediction error and smallest RMSE and $R^2$ error spread compared to BPWDNN, GDBPNN and SHLFFNN models have largest RMSE and $R^2$ error spread

The RMSE and $R^2$ values for the five different neural network models for prediction of antitubercular activity are comparable as shown in Figure 4. QRNN and GDBPMNN models appear to have slightly smaller RMSE and $R^2$ spreads than BPWDNN model. SHLFFNN and GDBPNN models appear to have larger RMSE and $R^2$ error spreads than BPWDNN model. Pair-wise comparisons of model RMSE and $R^2$ values using Student's $t$-test reveal that there is statistical difference in the prediction accuracies of the five neural network models. These results are shown in Table 1, which gives both the $p$-values and the absolute differences in RMSE and $R^2$ for the model comparisons. None of the $p$-values are smaller than the specified significance level α = 0.05. The null hypothesis is not rejected; in the context of this data set, there is no statistically significant difference in performance among these five neural network methods.

Table 1. Pair-wise comparisons of RMSE and $R^2$ differences and $p$-values

| **RMSE differences (upper diagonal) and $p$-values (lower diagonal)** | | | | | |
|---|---|---|---|---|---|
| | GDBPNN | SHLFFNN | GDBPMNN | BPWDNN | QRNN |
| GDBPNN | | -0.02361 | 0.14891 | 0.08679 | 0.20649 |
| SHLFFNN | 1.00000 | | 0.17252 | 0.11040 | 0.23011 |
| GDBPMNN | 0.49971 | 0.01617 | | -0.06212 | 0.05759 |
| BPWDNN | 1.00000 | 0.29577 | 1.00000 | | 0.11971 |
| QRNN | 0.08053 | 0.01048 | 1.00000 | 0.19054 | |
| **$R^2$ differences (upper diagonal) and $p$-values (lower diagonal)** | | | | | |
| | GDBPNN | SHLFFNN | GDBPMNN | BPWDNN | QRNN |
| GDBPNN | | -0.10573 | -0.23230 | -0.15382 | -0.26385 |
| SHLFFNN | 1.0000 | | -0.12657 | -0.04809 | -0.15812 |
| GDBPMNN | 0.5848 | 0.6052 | | 0.07848 | -0.03155 |
| BPWDNN | 1.0000 | 1.0000 | 1.0000 | | -0.11003 |
| Ridge | 0.3109 | 0.2511 | 1.0000 | 0.4586 | |

It should be observed that the $p$-value for this pair-wise comparison is 0.08053 and 0.3109 (Table 1) for RMSE and $R^2$, which is not valid at α = 0.05, but it is still a much smaller $p$-value than those obtained for the other four pair-wise comparisons. To test for pair-wise differences, we use Tukey differences.

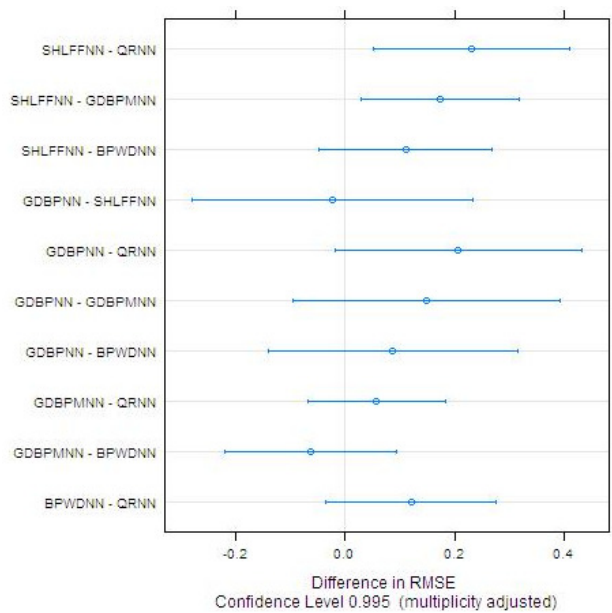

Figure 5. Asymptotic simultaneous confidence sets for Tukey all-pair neural network models comparisons of the RMSE errors after alignment.

As a major advantage compared to the non-parametric methods, can calculate simultaneous confidence intervals. Figure 5 shows the corresponding 99% model wise confidence intervals where the bars of a given comparison are outside the 0 difference in RMSE line there is a statistically meaningful difference at the 99% level present. The blue dot indicates the estimated magnitude of this difference. The differences between (GDBPNN,SHLFFNN) and (GDBPMNN,BPWDNN) are not significant, the corresponding confidence intervals intersect zero and overlap each other.

Table 2. Accuracy of predictions of the five neural network models

|  | RMSE | | $R^2$ | | MAE | | MPE | | RSE | |
|---|---|---|---|---|---|---|---|---|---|---|
|  | Train | Test | Train | Test | Train | Test | Train | Test | Train | Test |
| SHLFFNN | 0.4968453 | 0.4671671 | 0.7678053 | 0.8220075 | 0.08374804 | 0.2129987 | 45.97885 | 98.64839 | 0.4001315 | 0.4328208 |
| GDBPNN | 0.07103242 | 0.6128721 | 0.9918216 | 0.6911511 | 0.0001307464 | 0.2493231 | 1.129225 | 63.89045 | 0.008178501 | 0.7449092 |
| GDBPMNN | 0.113598 | 0.4118602 | 0.9878081 | 0.7096731 | 0.0722526 | 0.1411691 | 2.706897 | 73.75242 | 0.02091711 | 0.3364056 |
| BPWDNN | 0.3097244 | 0.3653125 | 0.8601063 | 0.7896301 | 0.0008788182 | 0.1641699 | 44.57782 | 69.02667 | 0.1554932 | 0.2646626 |
| QRNN | 3.710198e-06 | 0.2080948 | 1 | 0.9287288 | 6.934688e-08 | 0.05425151 | 0.0003052271 | 56.88983 | 2.231283e-11 | 0.08587886 |

In our study, neural network methods to predict antitubercular activity of Oxazolines and Oxazoles derivatives. In this case descriptor dataset is splits into training set and test set. Training set comprises seventy six observations and test set comprises twenty four observations.

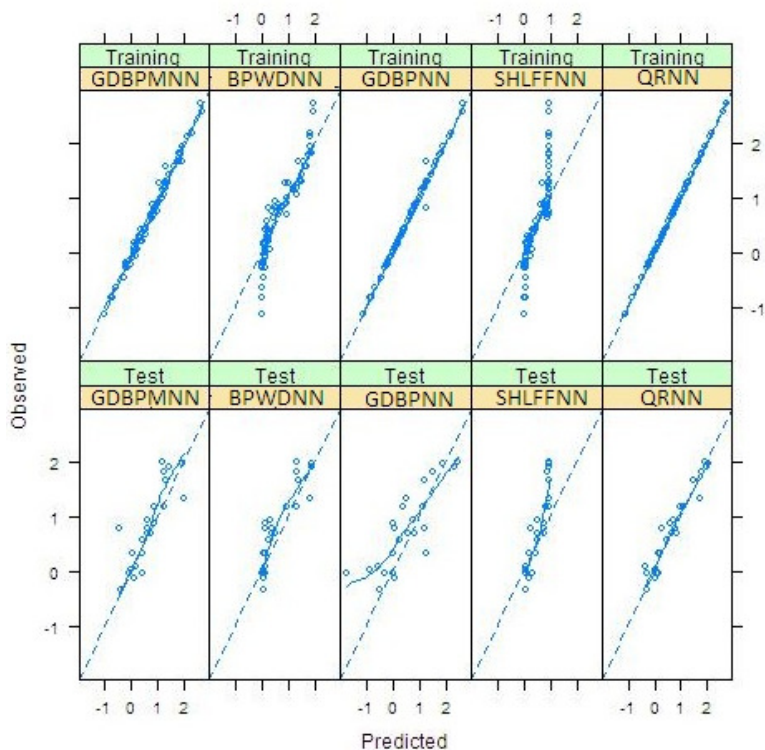

Figure 6. Comparison of prediction performance of trained and tested models obtained by five neural network methods for Oxazolines and Oxazoles derivatives descriptor dataset.

Predictive accuracy of all five neural network model evaluated as the coefficient determination i.e. $R^2$, root mean squared error (RMSE), mean absolute error(MAE), mean percentage error (MPE) and relative squared error(RSE). RMSE and $R^2$ provides baseline measures of predictive accuracy. All results reported are for the training set and test set. The predictive estimation results are summarized in Table 2.

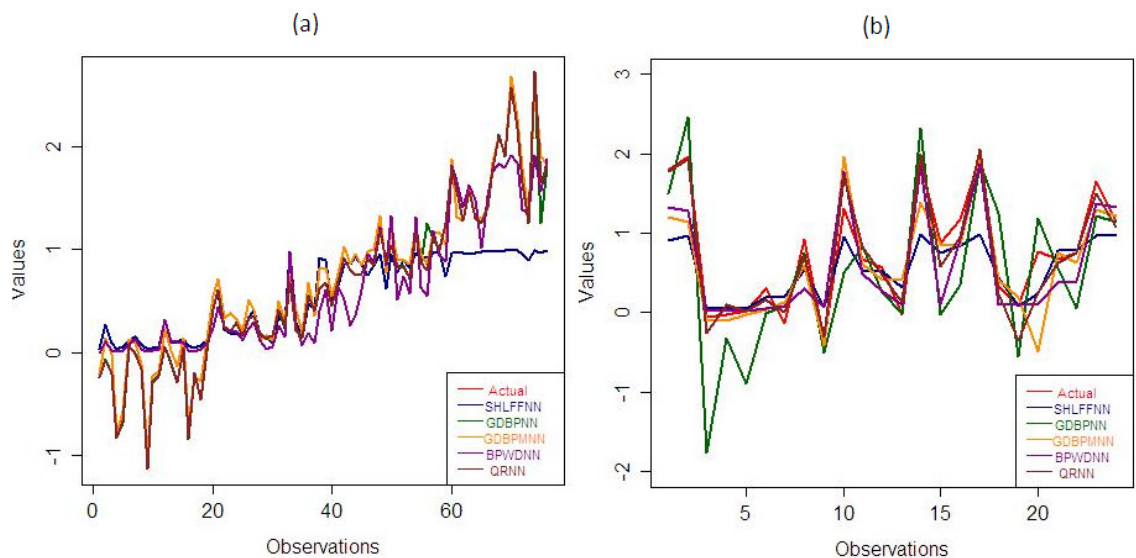

The obtained both RMSE and $R^2$ values of trained QRNN model are 3.710198e-06 and 1 are highly significant as well as RMSE and $R^2$ values of tested QRNN models are 0.2080948 and 0.9287288 are outstanding. The RMSE and $R^2$ values of trained GDBPNN model is 0.07103242 and 0.9918216 comparatively better than GDBPMNN model i.e. 0.113598 and 0.9878081. The RMSE and $R^2$ values of tested models are SHLFNN and BPWDNN are comparatively better than GDBPNN and GDBPMNN models. Figure 6 shows the performance comparison of the five methods for antitubercular activity prediction. In order to test and validate the five neural network models, the rest of five statistical tests are (MAE, MPE and RSE). These statistical tests signifies the QRNN model shows more significant than other four neural network models. These statistical tests are summarized in Table 2. Figure 7 shows the deviation from actual data of the five neural network models for trained and test datasets. Zero deviation shows that actual data and predicted QRNN model is overlapped is shown in Figure 7a. It means that predicted data has zero errors.

## 4. CONCLUSIONS

At first, important properties of neural network modelling methods and basic concepts of this were introduced. This technique is historically based on the attempt to model the way a biological brain processes the data. This study evaluated the ability of a five neural network models to predict antitubercular activity of Oxazolines and Oxazoles derivatives. we presented exploratory and inferential analyses of benchmark experiments. Benchmark experiments show that this method is the primary choice to evaluate neural network models. . It should be observed that the scheme can be utilized to compare a set of neural network techniques but does not offer a neural network model selection. The results for the non linear neural network models suggest that we may detect performance differences with fairly high power. We have compared the predictive accuracies with all five neural network models among QRNN model is outperformed overall predictive performance.

## ACKNOLDGEMENTS

We thankful to the Department of Computer Science Mangalore University, Mangalore India for technical support of this research.

**Authors**

**Doreswamy** received B.Sc degree in Computer Science and M.Sc Degree in Computer Science from University of Mysore in 1993 and 1995 respectively. Ph.D degree in Computer Science from Mangalore University in the year 2007. After completion of his Post-Graduation Degree, he subsequently joined and served as Lecturer in Computer Science at St. Joseph's College, Bangalore from 1996-1999.Then he has elevated to the position Reader in Computer Science at Mangalore University in year 2003. He was the Chairman of the Department of Post-Graduate Studies and research in computer science from 2003-2005 and from 2009-2008 and served at varies capacities in Mangalore University at present 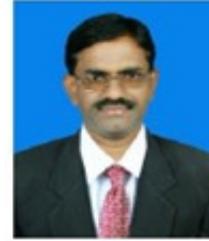
he is the Chairman of Board of Studies and Professor in Computer Science of Mangalore University. His areas of Research interests include Data Mining and Knowledge Discovery, Artificial Intelligence and Expert Systems , Bioinformatics ,Molecular modelling and simulation ,Computational Intelligence ,Nanotechnology, Image Processing and Pattern recognition. He has been granted a Major Research project entitled "Scientific Knowledge Discovery Systems (SKDS) for Advanced Engineering Materials Design Applications" from the funding agency University Grant Commission, New Delhi , India. He has been published about 30 contributed peer reviewed Papers at national/International Journal and Conferences. He received SHIKSHA RATTAN PURASKAR for his outstanding achievements in the year 2009 and RASTRIYA VIDYA SARASWATHI AWARD for outstanding achievement in chosen field of activity in the year 2010.

**Chanabasayya .M. Vastrad** received B.E. degree and M.Tech. degree in the year 2001 and 2006 respectively. Currently working towards his Ph.D Degree in Computer Science and Technology under the guidance of Dr. Doreswamy in the Department of Post-Graduate Studies and Research in Computer Science , Mangalore University
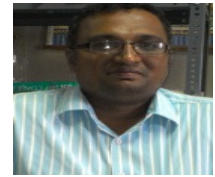